\documentclass[12pt]{article}
\usepackage{graphicx}

\def\Re{{\cal R \mskip-4mu \lower.1ex \hbox{\it e}\,}}
\def\Im{{\cal I \mskip-5mu \lower.1ex \hbox{\it m}\,}}
\def\ie{{\it i.e.}}
\def\eg{{\it e.g.}}

\def\etal{{\it et al.}}

\def\sub#1{_{\lower.25ex\hbox{$\scriptstyle#1$}}}
\def\tev{\,{\ifmmode\mathrm {TeV}\else TeV\fi}}
\def\gev{\,{\ifmmode\mathrm {GeV}\else GeV\fi}}
\def\mev{\,{\ifmmode\mathrm {MeV}\else MeV\fi}}
\def\mpl{\ifmmode M_{pl}\else $M_{pl}$\fi}
\def\mpl{\ifmmode \overline M_{Pl}\else $\bar M_{Pl}$\fi}
\def\to{\rightarrow}

\def\subw{_{\rm w}}
\def\mh{\ifmmode m\sbl H \else $m\sbl H$\fi}
\def\mch{\ifmmode m_{H^\pm} \else $m_{H^\pm}$\fi}
\def\mt{\ifmmode m_t\else $m_t$\fi}
\def\mc{\ifmmode m_c\else $m_c$\fi}
\def\mz{\ifmmode M_Z\else $M_Z$\fi}
\def\mw{\ifmmode M_W\else $M_W$\fi}
\def\mws{\ifmmode M_W^2 \else $M_W^2$\fi}
\def\mhs{\ifmmode m_H^2 \else $m_H^2$\fi}   
\def\mzs{\ifmmode M_Z^2 \else $M_Z^2$\fi}
\def\mts{\ifmmode m_t^2 \else $m_t^2$\fi}
\def\mcs{\ifmmode m_c^2 \else $m_c^2$\fi}
\def\mchs{\ifmmode m_{H^\pm}^2 \else $m_{H^\pm}^2$\fi}
\def\ztwo{\ifmmode Z_2\else $Z_2$\fi}
\def\zone{\ifmmode Z_1\else $Z_1$\fi}
\def\mtwo{\ifmmode M_2\else $M_2$\fi}
\def\mone{\ifmmode M_1\else $M_1$\fi}
\def\tb{\ifmmode \tan\beta \else $\tan\beta$\fi}
\def\xw{\ifmmode x\subw\else $x\subw$\fi}
\def\ch{\ifmmode H^\pm \else $H^\pm$\fi}
\def\lum{\ifmmode {\cal L}\else ${\cal L}$\fi}
\def\inpb{\,{\ifmmode {\mathrm {pb}}^{-1}\else ${\mathrm {pb}}^{-1}$\fi}}
\def\infb{\,{\ifmmode {\mathrm {fb}}^{-1}\else ${\mathrm {fb}}^{-1}$\fi}}
\def\epem{\ifmmode e^+e^-\else $e^+e^-$\fi}
\def\ppb{\ifmmode \bar pp\else $\bar pp$\fi}
\def\bsg{\ifmmode B\to X_s\gamma\else $B\to X_s\gamma$\fi}
\def\bsll{\ifmmode B\to X_s\ell^+\ell^-\else $B\to X_s\ell^+\ell^-$\fi}
\def\bstt{\ifmmode B\to X_s\tau^+\tau^-\else $B\to X_s\tau^+\tau^-$\fi}
\def\lamt{\ifmmode \tilde\lambda\else $\tilde\lambda$\fi}
\def\shat{\ifmmode \hat s\else $\hat s$\fi}
\def\that{\ifmmode \hat t\else $\hat t$\fi}
\def\uhat{\ifmmode \hat u\else $\hat u$\fi}

\newskip\zatskip \zatskip=0pt plus0pt minus0pt
\def\matth{\mathsurround=0pt}
\def\lsim{\mathrel{\mathpalette\atversim<}}

\def\atversim#1#2{\lower0.7ex\vbox{\baselineskip\zatskip\lineskip\zatskip
  \lineskiplimit 0pt\ialign{$\matth#1\hfil##\hfil$\crcr#2\crcr\sim\crcr}}}

\def\grtsim{\,\,\rlap{\raise 3pt\hbox{$>$}}{\lower 3pt\hbox{$\sim$}}\,\,}
\def\lsim{\,\,\rlap{\raise 3pt\hbox{$<$}}{\lower 3pt\hbox{$\sim$}}\,\,}

\renewcommand{\thefootnote}{\fnsymbol{footnote}}

\begin{document} \begin{titlepage}
\rightline{\vbox{\halign{&#\hfil\cr
&SLAC-PUB-14280\cr
}}}
\begin{center}
\thispagestyle{empty} \flushbottom { {\Large\bf Transverse Beam Polarization as an Alternate View into New Physics at CLIC  
\footnote{Work supported in part by the Department of Energy, Contract DE-AC02-76SF00515}
\footnote{e-mail:rizzo@slac.stanford.edu}}}
\medskip
\end{center}

\centerline{Thomas G. Rizzo}
\vspace{8pt} 
\centerline{\it SLAC National
Accelerator Laboratory, 2575 Sand Hill Rd., Menlo Park, CA, 94025}

\vspace*{0.3cm}

\begin{abstract}
In $e^+e^-$ collisions, transverse beam polarization can be a useful tool in studying the properties of particles associated with new physics 
beyond the Standard Model(SM). However, unlike in the case of measurements associated with longitudinal polarization, the formation of 
azimuthal asymmetries used to probe this physics in the case of transverse polarization requires both $e^\pm$ beams to be simultaneously 
polarized. In this paper we discuss the further use of transverse polarization as a  probe of new physics models at a high energy, $\sqrt s=3$ 
TeV version of CLIC. In particular, we show ($i$) how measurements of the sign of these asymmetries is sufficient to discriminate the 
production of spin-0 supersymmetric states from the spin-1/2 Kaluza-Klein excitations of Universal Extra Dimensions. Simultaneously, the 
contribution to this asymmetry arising from the potentially large SM $W^+W^-$ background can be made negligibly small. We then show ($ii$) 
how measurements of such asymmetries and their associated angular distributions on the peak of a new resonant $Z'$-like state can be used to 
extract precision information on the $Z'$ couplings to the SM fermions.
\end{abstract}



\renewcommand{\thefootnote}{\arabic{footnote}} \end{titlepage} 

%
%
%
%
%

\section{Introduction and Background}

With the LHC now running at $\sqrt s=7$ TeV and eventually running at 14 TeV, New Physics(NP) beyond the SM at the TeV scale is expected 
to show up sometime soon. Based on theoretical prejudice this NP should assist in our understanding of the hierarchy problem and may be intimately 
involved in the breaking of electroweak symmetry. Although there are many speculations, it is not known what form this new physics might 
take with Supersymmetry{\cite {susy}}, extra dimensions{\cite {ed}} and extended gauge theories{\cite {zp}} being among the potential  
candidates. Once the LHC discovers this physics the more difficult task of uncovering the underlying theory will still lay ahead. 
In all probability it is quite likely that the LHC may not provide us with sufficient information to address this problem in full detail and many in 
the community expect that the precision measurements available at a lepton collider, whose center of mass energy will depend on the precise 
mass scale of this NP, will be necessary to provide the complete answer to these questions.

After the determination of its mass, the most elementary and important properties of any new particle are its spin and the nature of its couplings 
to the familiar SM fields; various tools will be necessary to obtain this information. The possibility of using transverse beam polarization in 
$e^+e^-$ collisions to explore the details of many various NP scenarios has now become rather well-established{\cite {huge}} over the last 
few years. However, unlike in the case of the physics studies employing longitudinal beam polarization, to study NP with asymmetries 
produced in the transversely polarized case both $e^\pm$ beams need to be polarized. The reason for this is that the corresponding asymmetry 
parameters, $A$, associated with azimuthal angular distributions are directly proportional to the product of these two polarizations, \ie, 
$A \sim p_1^Tp_2^T$. {\footnote {Here we also recall that transverse beam 
polarization is the `natural' state of polarized beams at such colliders and that spin rotators are employed to produce the more conventionally 
studied case of longitudinal beam polarization.}} In this paper we will explore two scenarios in which transverse polarization can be used to 
obtain useful information about the properties of new particles in high energy $e^+e^-$ collisions. To be specific, we will focus on very high 
energy collisions, $\sqrt s=3$ TeV, as are eventually envisioned at CLIC. We will assume that integrated luminosities in the few 
$ab^{-1}$ range are available and further assume that that the product of the magnitudes of the transverse polarizations $|P_1^TP_2^T|=0.5$, 
not an unreasonable value given the estimates for possible positron 
polarization, for purposes of demonstration. Our philosophy in the preliminary studies presented here is {\it not} to ignore the well-known 
techniques that employ longitudinal polarization to address these issues but to explore instead the capabilities of transverse polarization to 
provide an alternative window into the same NP.

The first scenario we consider is new particle spin discrimination. As is well-known, measurements at 
lepton colliders provide at least two relatively straightforward ways to determine the spin of a particle which is pair produced in, \eg, the 
$s$-channel. These techniques would then allow us to discriminate, \eg, the production of SUSY particles from the Kaluza-Klein(KK) excitations 
of Universal Extra Dimensions (UED){\cite {UED}}. First, a threshold scan would reveal a 
slow $\sim \beta^3$ turn-on in the case of spin-0 sleptons and squarks, with 
$\beta$ being the particle's velocity,  whereas the spin-1/2 KK excitations would instead have a $\sim \beta$ behavior. Second, far above 
the threshold region, the production of the spin-0 SUSY states would lead to a $\sim 1-\cos^2\theta$ angular distribution 
whereas the KK states, being vector-like, would 
instead yield a $\sim 1+\cos^2\theta$ distribution. In a classic work{\cite {Marco}}, it has been explicitly shown how such observables can be used at 
CLIC, running at $\sqrt s$=3 TeV, to differentiate $\sim 500$ GeV smuons from the level-1 KK muons in UED. Although these two techniques are 
powerful there could be situations where even observing these states associated with NP might prove to be difficult{\cite {Berger:2007ut,
Berger:2007yu}} so that it can be worthwhile having an additional tool at hand to assist in spin discrimination. As we will demonstrate 
below, transverse polarization can provide such an additional tool. To be specific, the analysis presented below is based on a few simple 
observations: ($i$) The magnitudes of the transverse polarization asymmetries, $A$, for scalars and vector-like fermions are large and are of 
opposite sign whereas $A$ for $W$-boson pair production, the largest SM background, is highly suppressed. ($ii$) The $\cos \theta$ dependence of 
$A$ for both scalars and vector-like fermions behaves as $\sim 1-\cos^2\theta$; this also remains true in the cases of $W$-pair production to a very 
good approximation so that the signal shape is not distorted by the background. ($iii$) Suitable angular cuts can be used to remove the bulk of this $W$-pair 
background.
  
In the second scenario we consider the existence of a $Z'$-like state and ask the hypothetical question: `What information about this state can be 
provided through the use of transverse polarization if it is employed instead of longitudinal polarization?'. As is very well-known, the existence 
of longitudinal polarization allows for the determination of the Left-Right polarization asymmetry, $A_{LR}$, as well as the polarized Forward-Backward 
asymmetry for any final state fermion $f$, $A_{FB}^{pol}(f)$, from which coupling information can be extracted when combined with unpolarized data such 
as partial widths. Here we will show that detailed measurements of the decay azimuthal angular distribution can be also employed to extract analogously 
useful information on the couplings of this $Z'$ state to the SM fermions.

\section{Spin Discrimination}

\begin{figure}[htbp]
\centerline{
\includegraphics[width=8.5cm,angle=90]{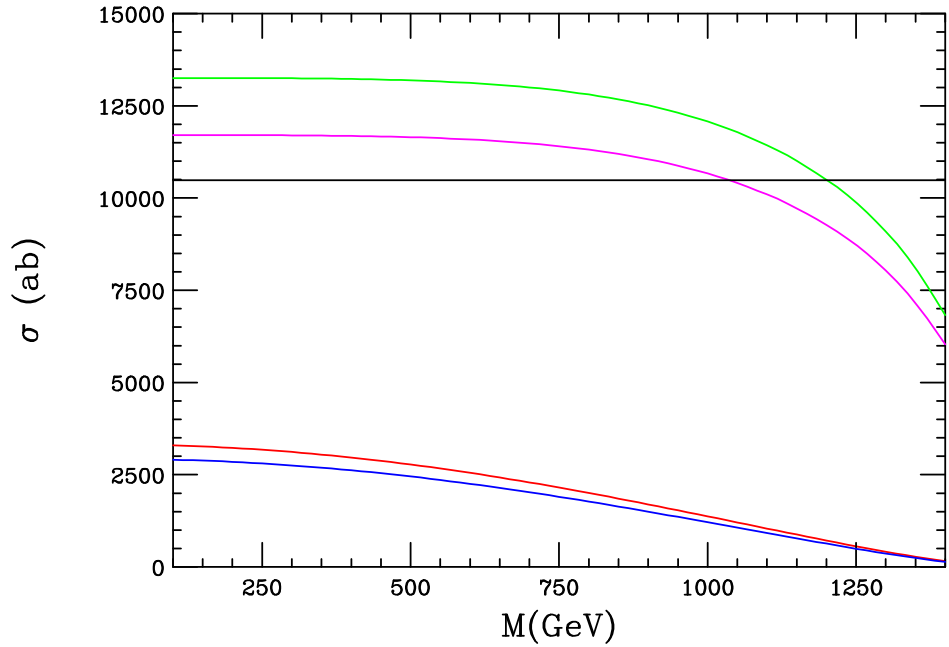}}
\vspace*{0.1cm}
\caption{Cross section for the production of left- and right-handed smuons (red and blue curves, respectively) and the corresponding level-1 KK 
states in UED (green and magenta curves, respectively) as a function of their mass at a 3 TeV $e^+e^-$ collider. The corresponding cross section for 
$W$-pair production followed by their subsequent decay to muons is also shown (solid line) for comparison purposes.}
\label{fig1}
\end{figure}

At CLIC luminosities ($\sim 1 ab^{-1}$) and energies ($\sim 3$ TeV), kinematically accessible SUSY particles and KK excitations in UED are expected to  
produce rather significant event samples for detailed study unless these states are accidentally close to threshold. Thus, statistics should not 
really be much of an issue in performing such analyses. For the specific cases of smuons and the level-1 KK muons in UED, these cross sections are 
shown in Fig.~\ref{fig1} together with that for $W$-pair production with both $W$'s decaying to muons which provides the largest SM background 
after $\gamma\gamma$ initiated states are removed with the appropriate cuts. (In what follows it will be assumed for simplicity that both the smuons and 
the KK states will decay with 100$\%$ branching fractions to muons plus missing energy.) 

Let us first consider the production of smuons; ignoring for the moment small terms proportional to $\sim \Gamma_Z/M_Z$ away from the $Z$-pole, 
the double differential cross section is given by 
\begin{equation}
{{d\sigma}\over {dz d\phi}}= {{\alpha^2 \beta^3}\over {8s}} (1-z^2) ~[F_A-P_1^T P_2^T F_B \cos 2\phi]\,,
\end{equation}
where $\beta$ is the speed of the smuon as usual, $z=\cos \theta$, and 
\begin{equation}
F_{A(B)}=Q_e^2Q_f^2+2Q_eQ_fv_eG_VR_1+(v_e^2\pm a_e^2)G_V^2R_2\,,
\end{equation}
with $v(a)_e$ being the vector(axial-vector) coupling of the electron to the $Z$, $Q_{e,f}$ being the electric charges of the electron and final state 
particle such that 
\begin{equation}
G_V=B(T_{3f}-x_wQ_f), ~~v_e=B(-1/4+x_w), ~~a_e=-B/4\,.
\end{equation}
for any $f$ in the final state. Here we have defined the familiar dimensionless quantities
\begin{equation}
R_1=s(s-m_Z^2)/[(s-m_Z^2)^2+\Gamma_Z^2M_Z^2], ~~R_2=s^2/[(s-m_Z^2)^2+\Gamma_Z^2M_Z^2]\,,
\end{equation}
with $B=[{{\sqrt 2 G_FM_Z^2}\over {\pi \alpha(M_Z)}}]^{1/2}$ employed in the coupling definitions above. We note that the $z-$ and $\phi-$dependencies of this 
distribution are observed to factorize in a rather simple manner. Integration over $z$ and normalizing to the total cross section yields the 
azimuthal distribution
\begin{equation}
{{1}\over {\sigma}} {{d\sigma}\over {d\phi}}= {{1}\over {2\pi}} (1+\lambda \cos 2\phi)\,,
\end{equation}
where we have defined 
\begin{equation}
\lambda =-P_1^T P_2^T {{F_B}\over {F_A}}\,,
\end{equation}
Further separate integration over the `odd' and `even' regions where $\cos 2\phi$ takes on opposite signs yields the transverse polarization asymmetry 
\begin{equation}
A= {{\int_{odd} d\sigma}\over {\int_{all} d\sigma}} = {{2\lambda}\over {\pi}}\,.
\end{equation}
If we had inverted the orders of the $z$ and $\phi$ integrations we would observe that this azimuthal asymmetry itself has a very simple $\sin^2 \theta$ 
behavior:
\begin{equation}
{{dA}\over {dz}}= {{3}\over {4}} (1-z^2) ~A\,. 
\end{equation}

Now let us examine the corresponding results for the case of the level-1 KK muons; recalling that these are vector-like fermions we quickly obtain 
\begin{equation}
{{d\sigma}\over {dz d\phi}}={{\alpha^2 \beta}\over {4s}} ~\Big(F_A[(1+z^2)+(1-\beta^2)(1-z^2)]+P_1^T P_2^T (1-z^2)\beta^2F_B \cos 2\phi \Big)\,,
\end{equation}
where all of the quantities are as defined above. Following the same procedure as in the case of smuons we find that for particles with the same electroweak 
quantum numbers 
\begin{equation}
A_{KK}= {{-\beta^2}\over {3-\beta^2}} ~A_{\tilde \mu}\,,
\end{equation}
with the asymmetry having same $\sin^2\theta$ angular dependence as was found for smuons. Thus both sets of particles yield asymmetries of comparable size 
but of opposite sign. 

In the case of $W$-pair production the analogous differential cross-section can be written as{\cite {fkj}} 
\begin{equation}
{{d\sigma}\over {dz d\phi}}\sim |M_L(z)|^2+|M_R(z))|^2+2P_1^T P_2^T Re(M_L(z)M_R(z)^*)\cos 2\phi \,,
\end{equation}
where $M_{L,R}$ are the corresponding, $z$-dependent, left- and right-handed helicity amplitudes. The resulting transverse polarization asymmetry is 
then found to be proportional to the ratio of the integrals over $z$:
\begin{equation}
A \sim {{\int dz~ 2Re\big(M_LM_R^* \big)}\over {\int dz~ \big(|M_L(z)|^2+|M_R(z))|^2\big)}}\,.
\end{equation}
Now as is well-known, in the case of $W$-pair production, since the $W$ couples in a purely left-handed manner to the SM fermions we can 
(symbolically) observe that $M_L >> M_R$. This leads to a rather small value for the asymmetry since $A\sim M_R/M_L$; to obtain the corresponding 
$A(z)$ distribution in this case, since it is defined relative to the total cross section,  we simply omit the $z$ integration in the numerator 
above. One then finds that $A(z)$ in the case of $W$-pair production also roughly behaves as $1-z^2$ at CLIC energies.
\begin{figure}[htbp]
\centerline{
\includegraphics[width=8.5cm,angle=90]{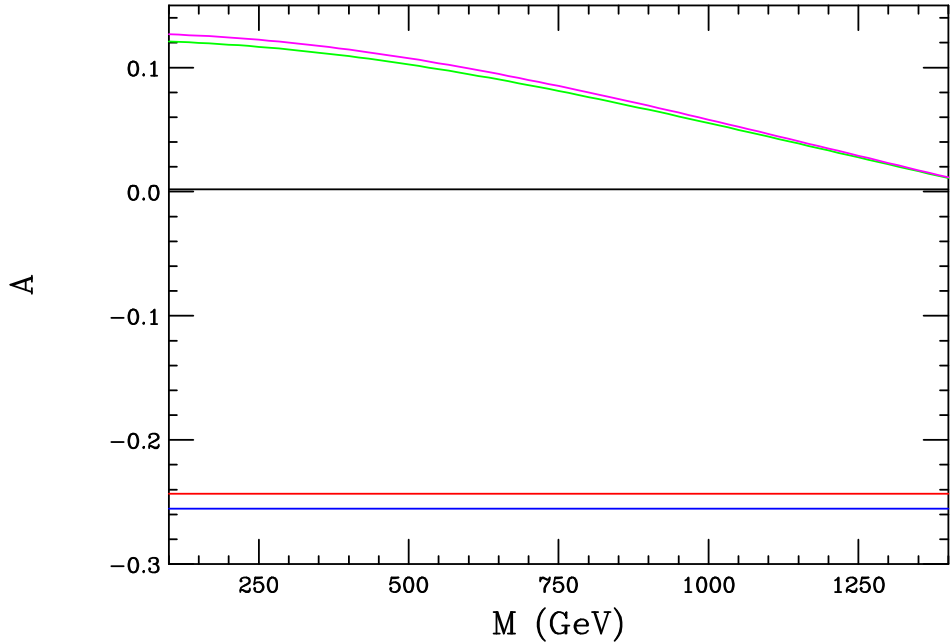}}
\vspace*{0.1cm}
\caption{The transverse polarization asymmetry, $A$, for smuons and level-1 KK muons as a function of their masses at a 3 TeV $e^+e^-$ collider 
with the curves labeled as in the previous figure.}
\label{fig2}
\end{figure}

\begin{figure}[htbp]
\centerline{
\includegraphics[width=7.5cm,angle=90]{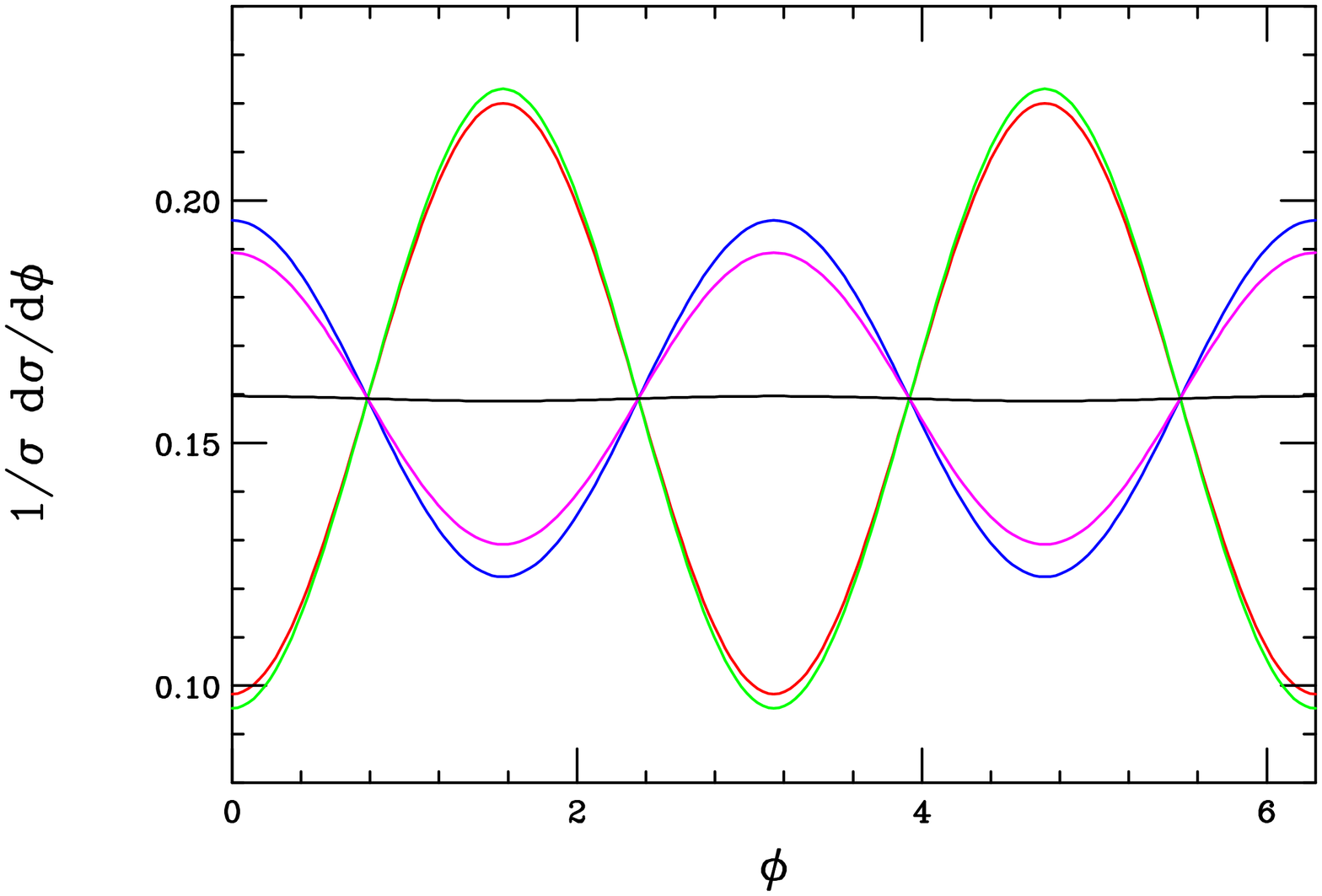}}
\vspace*{0.1cm}
\centerline{
\includegraphics[width=7.5cm,angle=90]{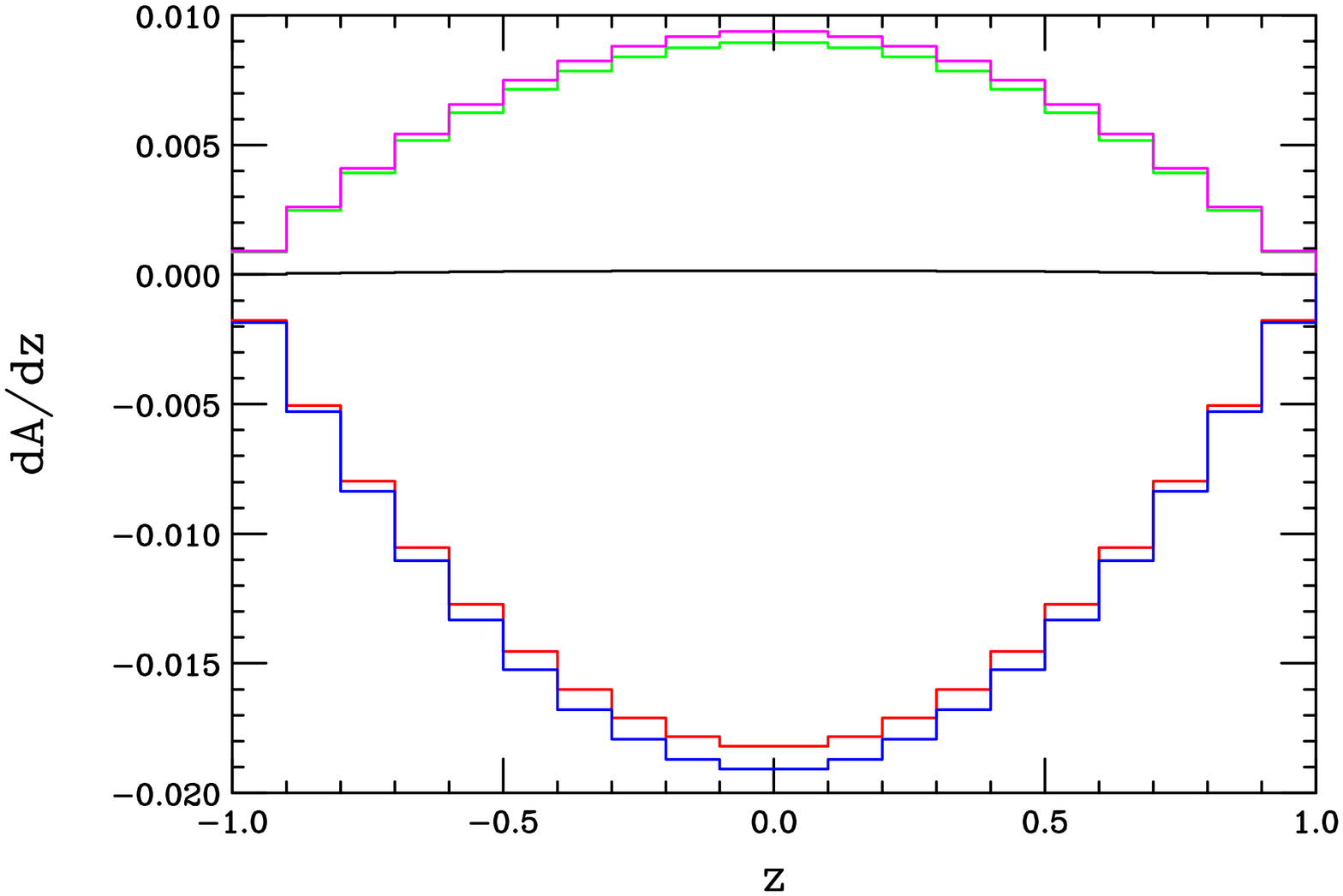}}
\vspace*{0.1cm}
\caption{(Top) The idealized azimuthal angular distributions for 500 GeV smuons and level-1 KK muons at a 3 TeV $e^+e^-$ collider. The results for $W$-pairs 
is also shown for comparison. The curves for smuons and muon KK states are interchanged in comparison to the previous figures. (Bottom) Binned $A(z)$ 
distributions for the same cases as described in the previous figure.}
\label{fig3}
\end{figure}

Fig.~\ref{fig2} shows the values of the transverse polarization asymmetry, $A$, for both smuons and level-1 KK muons, as a function of their mass, as 
well as for $W$-pairs at a 3 TeV CLIC. Whereas 
both smuons and muon KK states are observed to have large transverse polarization asymmetries, that for $W$'s is seen to be highly suppressed by 
comparison due to the dominantly LH couplings of the $W$. The upper panel of Fig.~\ref{fig3} shows the corresponding shapes of the idealized azimuthal angular 
distributions for these same three states whereas in the lower panel of Fig.~\ref{fig3} we find their binned $A(z)$ distributions. Since the $W$-pair 
asymmetry is so small in comparison to the two cases of interest it is difficult to see that it also has a roughly $\sin ^2 \theta$ shape. 

\begin{figure}[htbp]
\centerline{
\includegraphics[width=7.5cm,angle=90]{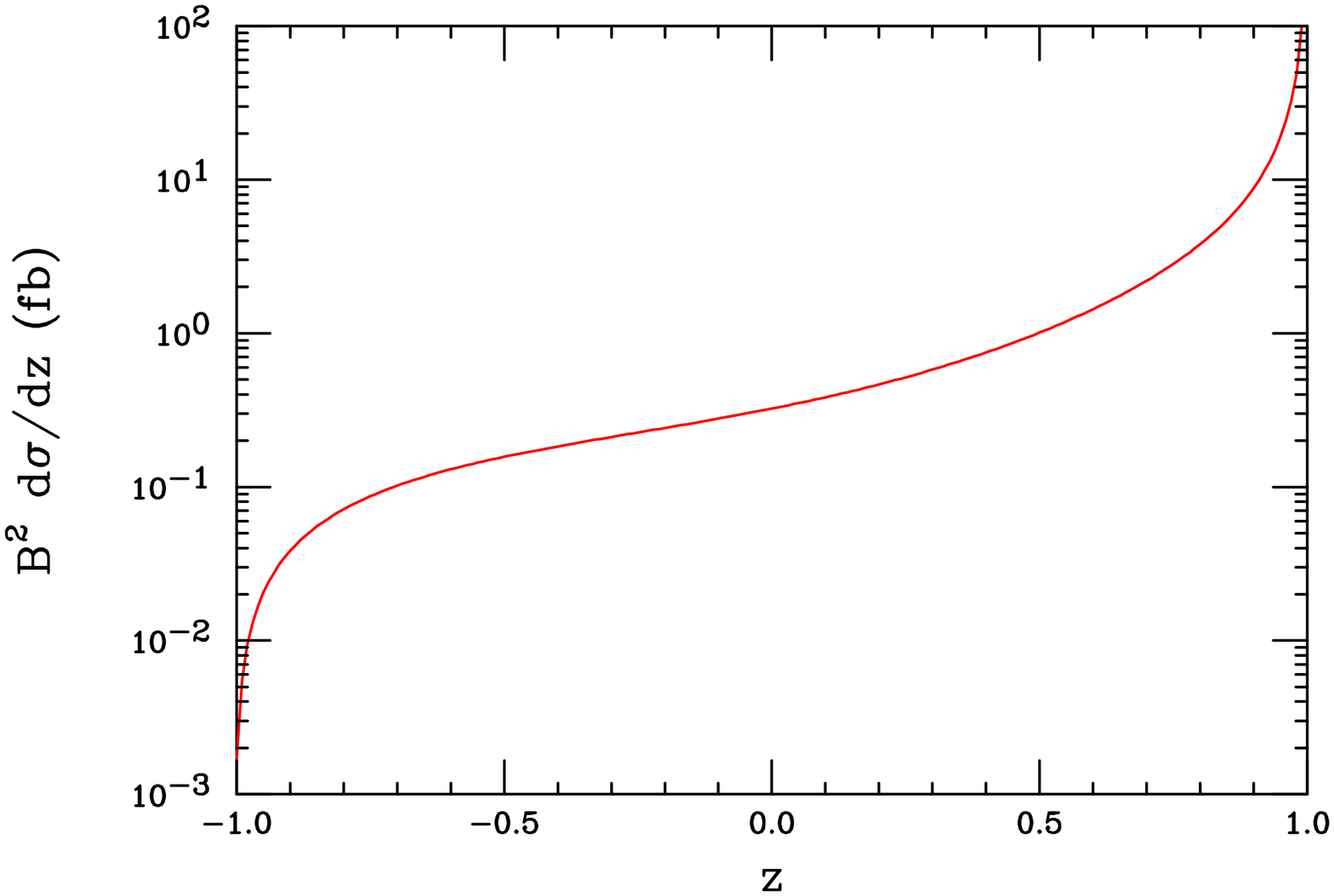}}
\vspace*{0.1cm}
\centerline{
\includegraphics[width=7.5cm,angle=90]{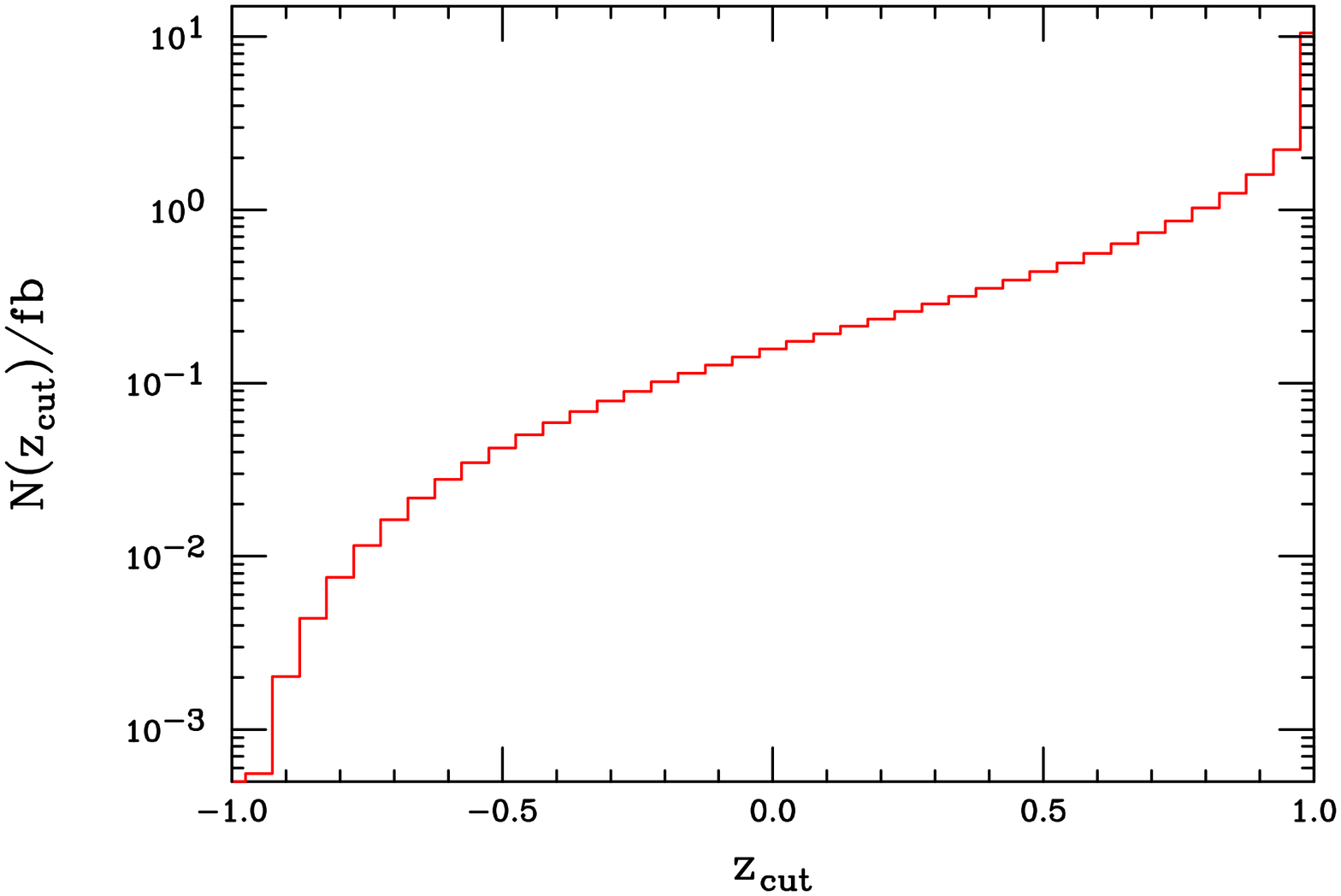}}
\vspace*{0.1cm}
\caption{(Top) Angular distribution for unpolarized $e^+e^-\to W^+W^-$ at $\sqrt s=$3 TeV. (Bottom) The number of $W$ induced muon events at this same energy 
requiring the negative muon to lie in the range $-1\leq z\leq z_{cut}$.}
\label{fig4}
\end{figure}

Although the $W$-pair background has a very small asymmetry, it's presence as a background will end up diluting the asymmetry signal from either of the 
two NP sources making 
spin discrimination more difficult. Ordinarily, when we are performing the identical parallel study employing longitudinal polarization, we can freely choose 
this polarization to be right-handed to remove a very large part of the $W$-pair induced background; we can't do that here so we need to resort to some 
other cut(s) to remove the $W$-pair contamination. We recall, however, that the $z$-distribution for $W$-pair production, which for 
$\sqrt s=3$ TeV is shown in the top panel of 
Fig.~\ref{fig4}, is highly peaked in the forward direction. This means that the negatively charged muon from the $W^-$ will be correspondingly very forwardly  
peaked due to the large boost. On the otherhand, negatively charged muons arising from either the decay of the smuons or KK states will be just as likely to 
go in either the forward or backward directions since the angular distributions shown above for the production of pairs of these particles are seen to be 
even functions of $z$. Thus removing events with negatively(positively) charged muons in the forward(backward) hemisphere will reduce the signal by only a 
factor of 2 while substantially reducing the $W$ induced background. In order too see how large of an effect this cut has on this background, the lower 
panel in Fig.~\ref{fig4} shows the $W$ angular event rate integrated over the range $-1 \leq z\leq z_{cut}$. For $z_{cut}=0$, \ie, performing the cut as 
described above and removing negative(positive) muons in the forward(backward) hemisphere, 
this is seen to reduce the background by more than a factor of $\simeq 60$ which is more than adequate for our purposes since the original 
$W$-pair cross section and the signal cross sections are roughly comparable. 

\begin{figure}[htbp]
\centerline{
\includegraphics[width=7.5cm,angle=90]{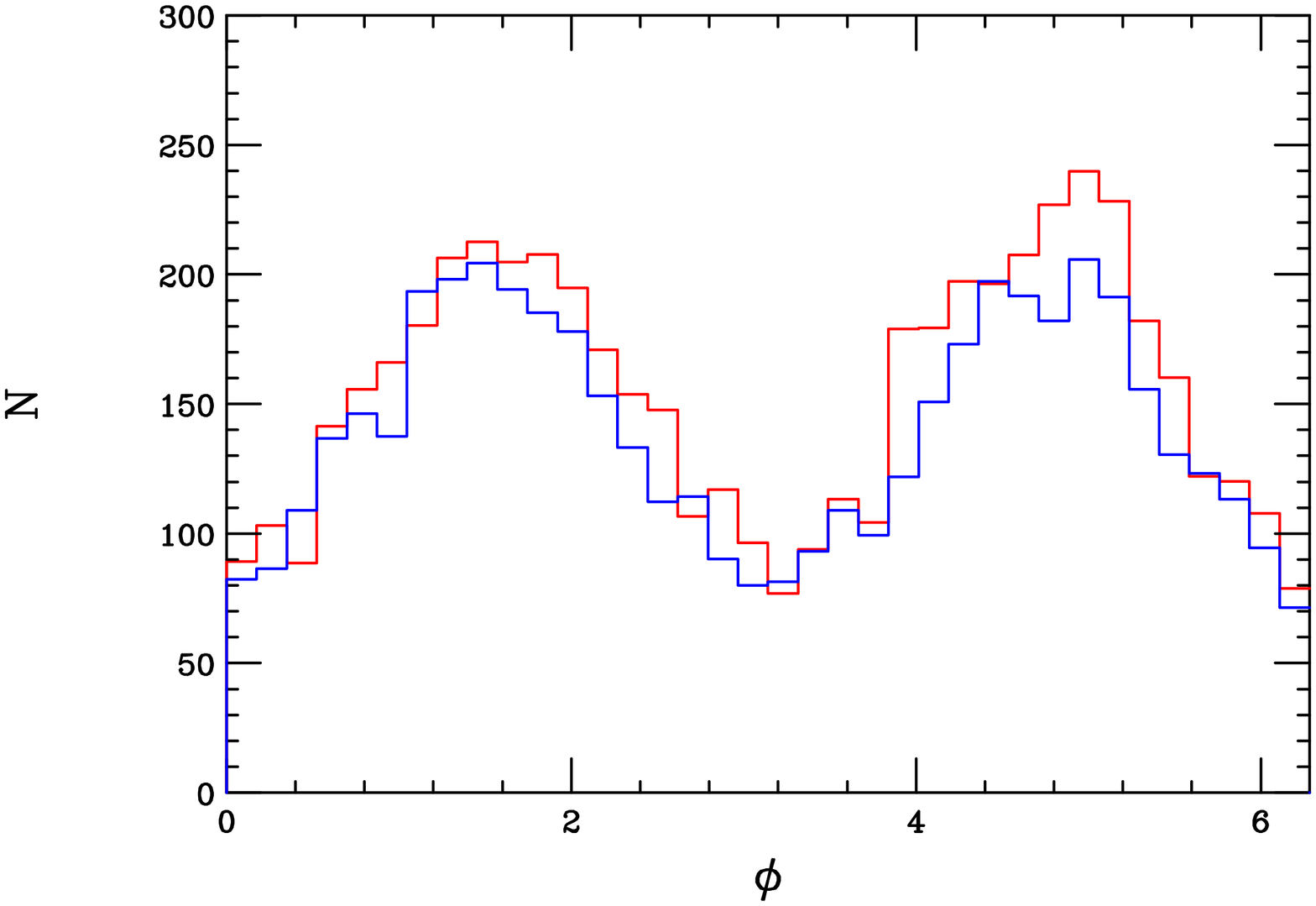}}
\vspace*{0.1cm}
\centerline{
\includegraphics[width=7.5cm,angle=90]{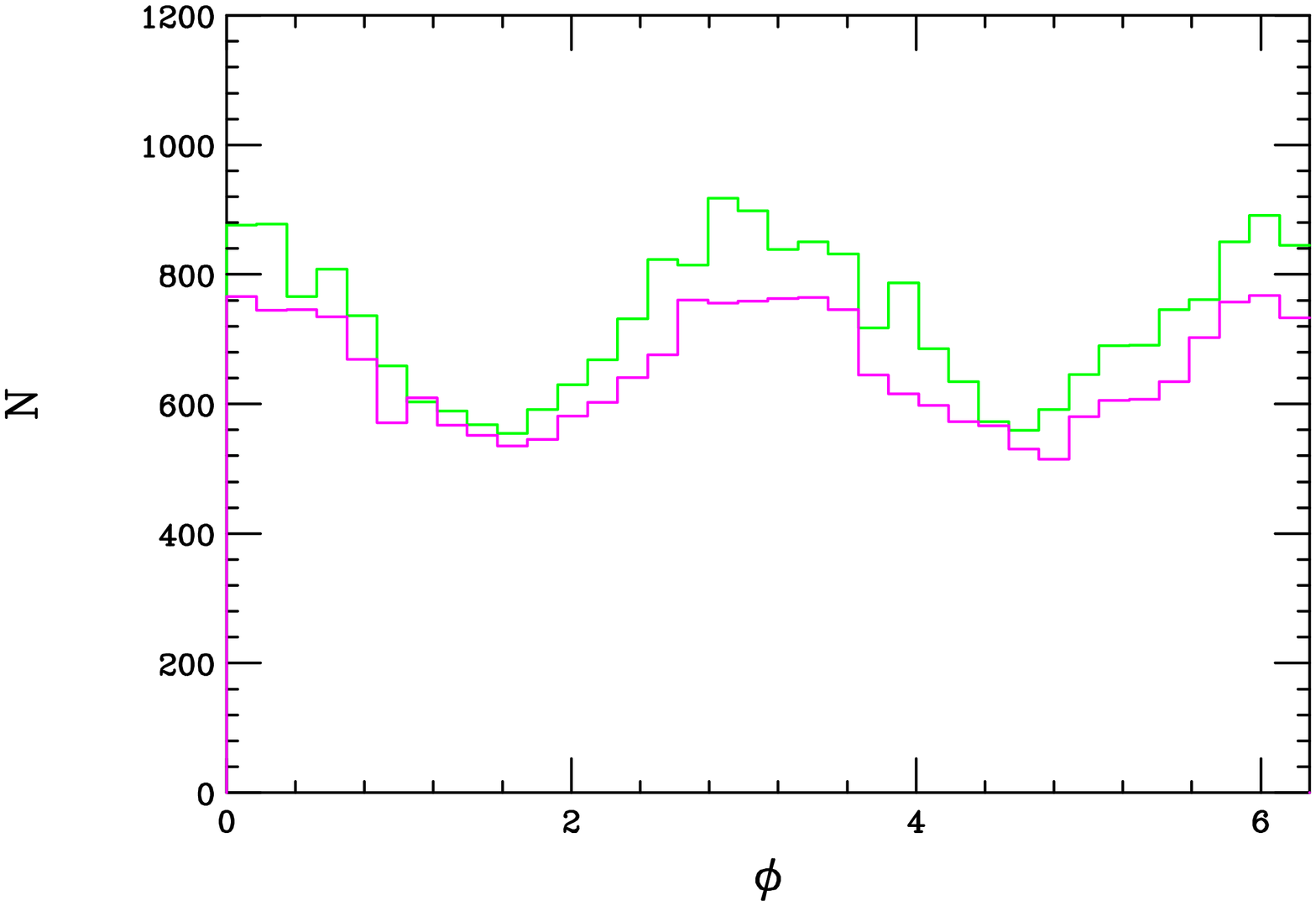}}
\vspace*{0.1cm}
\caption{Azimuthal event distributions for smuons(top) and KK muons(bottom) assuming masses of 500 GeV, $\sqrt s=3$ TeV and a luminosity of 2 $ab^{-1}$. 
The histogram color labels are as in the previous figures.}
\label{fig5}
\end{figure}

\begin{figure}[htbp]
\centerline{
\includegraphics[width=8.5cm,angle=90]{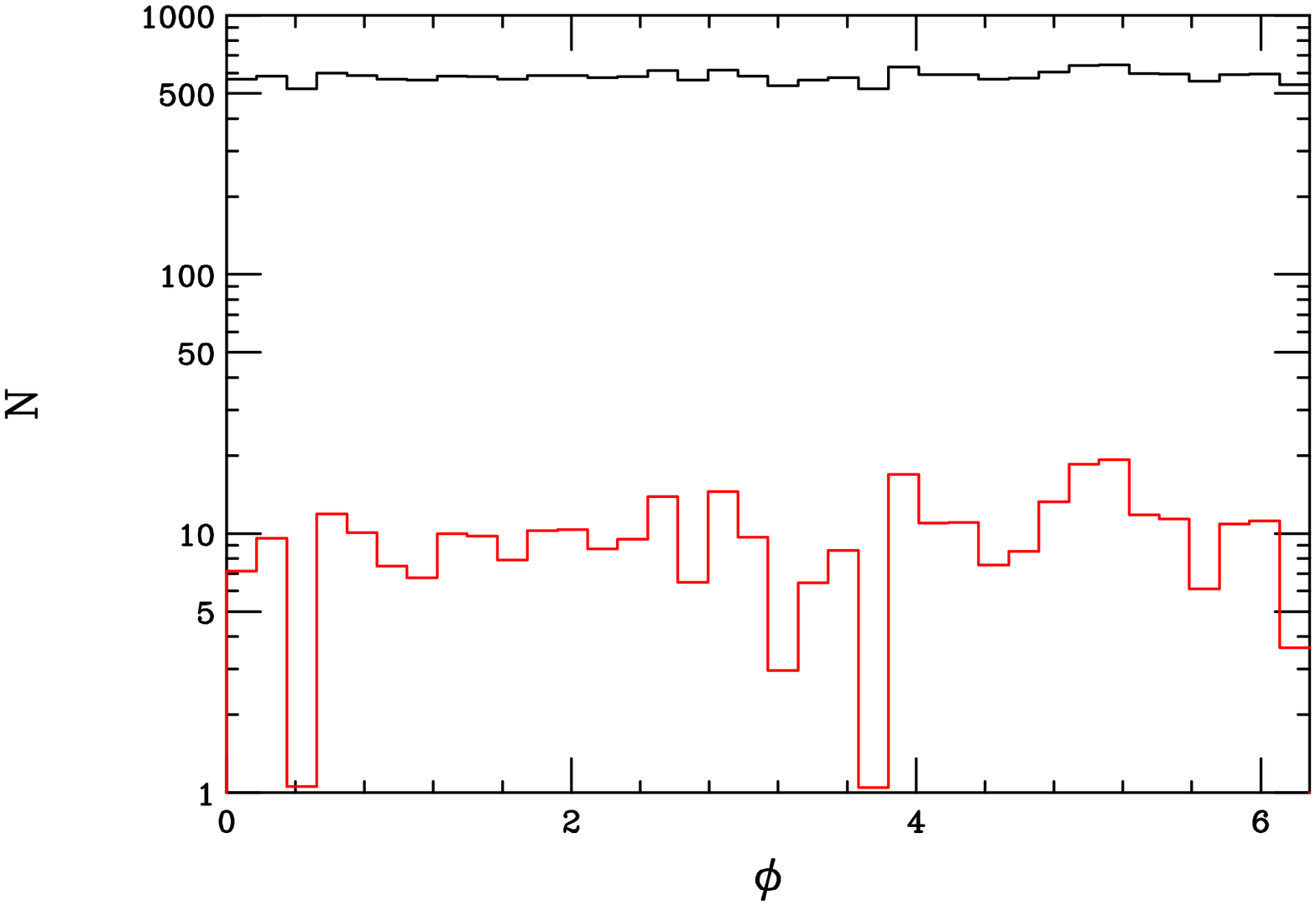}}
\vspace*{0.1cm}
\caption{Same as the previous figure but now for the $W$-pair background both before(black histogram) and after(red histogram) the cut on $z$ is applied.}
\label{fig6}
\end{figure}

What will these azimuthal angular distributions for the NP look like at CLIC?  In order to be specific, we assume smuon/KK masses of 500 GeV and an integrated 
luminosity of 2 $ab^{-1}$; the results for the event distributions are then shown in Fig.~\ref{fig5}. The corresponding result for the $W$-induced background, 
both before and after applying the angular cut discussed above, is shown in Fig.~\ref{fig6}; here we see that the cut makes this background negligible. We note 
that the azimuthal distributions for the smuons and the KK states are quite easily distinguishable for these masses with the assumed polarizations and integrated 
luminosities. 

This preliminary analysis indicates that transverse polarization asymmetries may be a useful tool at CLIC to help to discriminate particle spins. A more 
detailed study including full SM backgrounds, ISR/beamstrahlung and detector effects should be performed to verify these results. 

\section{New $Z'$ Coupling Determinations}

If a new $Z'$-like resonance is discovered at the LHC, we will want to know all of its properties, in particular its couplings to the SM fields, as well 
as the underlying theory which gave rise to it. Unfortunately, it is very likely that the LHC will be unable to perform this analysis in all generality, even 
if the $Z'$ is relatively light, due to the lack of a sufficient number of observables and limited integrated luminosity{\cite {zp,frank}}. If this 
is indeed the case, 
then the data from an $e^+e^-$ collider will be crucial, especially so if the $Z'$ mass is within the $\sqrt s$ range of this collider so that we can then sit 
on the $Z'$ pole. Here we imagine that such a state exists with a mass of 3 TeV (or less) and that we can sit on top of this resonance with CLIC. 

Ordinarily, with longitudinal beam polarization, the following observables are available to perform coupling determinations for any final state $f$: 
$\Gamma_f$, the partial widths, $A_{FB}(f)$, the Forward-Backward 
asymmetries, $A_{FB}^{pol}(f)$, the Polarized Forward-Backward asymmetries and $A_{LR}$, the Left-Right asymmetry. If longitudinal polarization is {\it not} 
available then we can't employ the last two observables and we need some `replacements' from obtained by the use of transverse polarization. 
To this end let us re-examine 
the normalized azimuthal angular distribution for massless fermions on top of the $Z'$ resonance; the general form of this distribution is given by 
\begin{equation}
{{1}\over {\Gamma}} {{d\Gamma}\over {d\phi}} \sim 1+{{1}\over {2}}P_1^T P_2^T\big(\lambda \cos 2\phi-\tau_f \sin 2\phi\big)\,,
\end{equation}
where the parameter $\lambda$, apart from a different choice of normalization factor, was described above. On the $Z'$ pole, we find that 
$\lambda$ only depends upon the vector 
and axial vector couplings of the electron to the $Z'$ and in that sense is completely analogous to $A_{LR}$: 
\begin{equation}
\lambda ={{v_e'^2-a_e'^2}\over {v_e'^2+a_e'^2}}\,.
\end{equation}
Here, and in what follows, a (un)primed coupling is one that corresponds to a coupling to the $(Z)Z'$. 
The parameters $\tau_f$ do not appear in the original expression for this distribution in the previous section since they originate from the absorptive part of 
the amplitude and are not significant away from resonances and were dropped in that analysis. On the $Z'$ pole, however, we find (dropping terms that are 
subleading in $M_Z^2/M_{Z'}^2 <<1$), that  
\begin{equation}
\tau_f =2~{{Q_eQ_fa_e'v_f'+v_ea_e'(v_fv_f'+a_fa_f')}\over {(v_e'^2+a_e'^2)(v_f'^2+a_f'^2)}}~{{\Gamma_{Z'}}\over {M_{Z'}}}\,,
\end{equation}
which are parametrically `small' since the width-to-mass ratio for a typical $Z'$ is likely to be in the range of a few percent. However, given the anticipated 
CLIC luminosity the event samples are expected to be huge as the peak cross section ({\it after} ISR and beamstrahlung corrections are accounted for) for the 
fermion $f$ is expected to be $\sigma_f \simeq 5.7 \times 10^6 (B_eB_f/0.01)$ ab where $B_{e(f)}$ are the electron and $f$ branching fractions of the $Z'$. 
With this enormous amount of statistics,  precision measurements of $\lambda$ and the $\tau_f$ should be rather straightforward and it is likely that many 
of these and other observables on the $Z'$ resonance will become systematics limited.

\begin{figure}[htbp]
\centerline{
\includegraphics[width=7.5cm,angle=90]{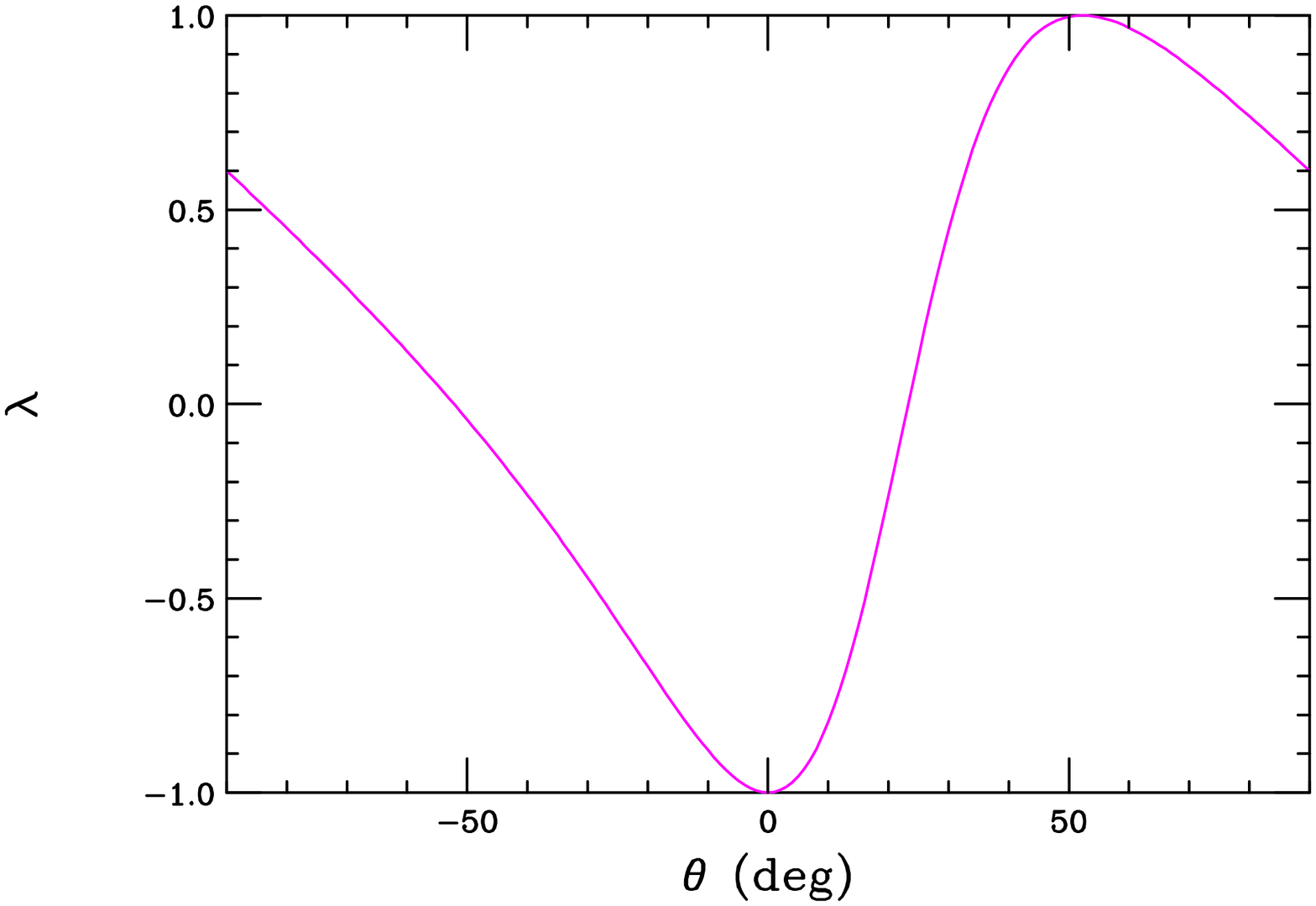}}
\vspace*{0.1cm}
\centerline{
\includegraphics[width=7.5cm,angle=90]{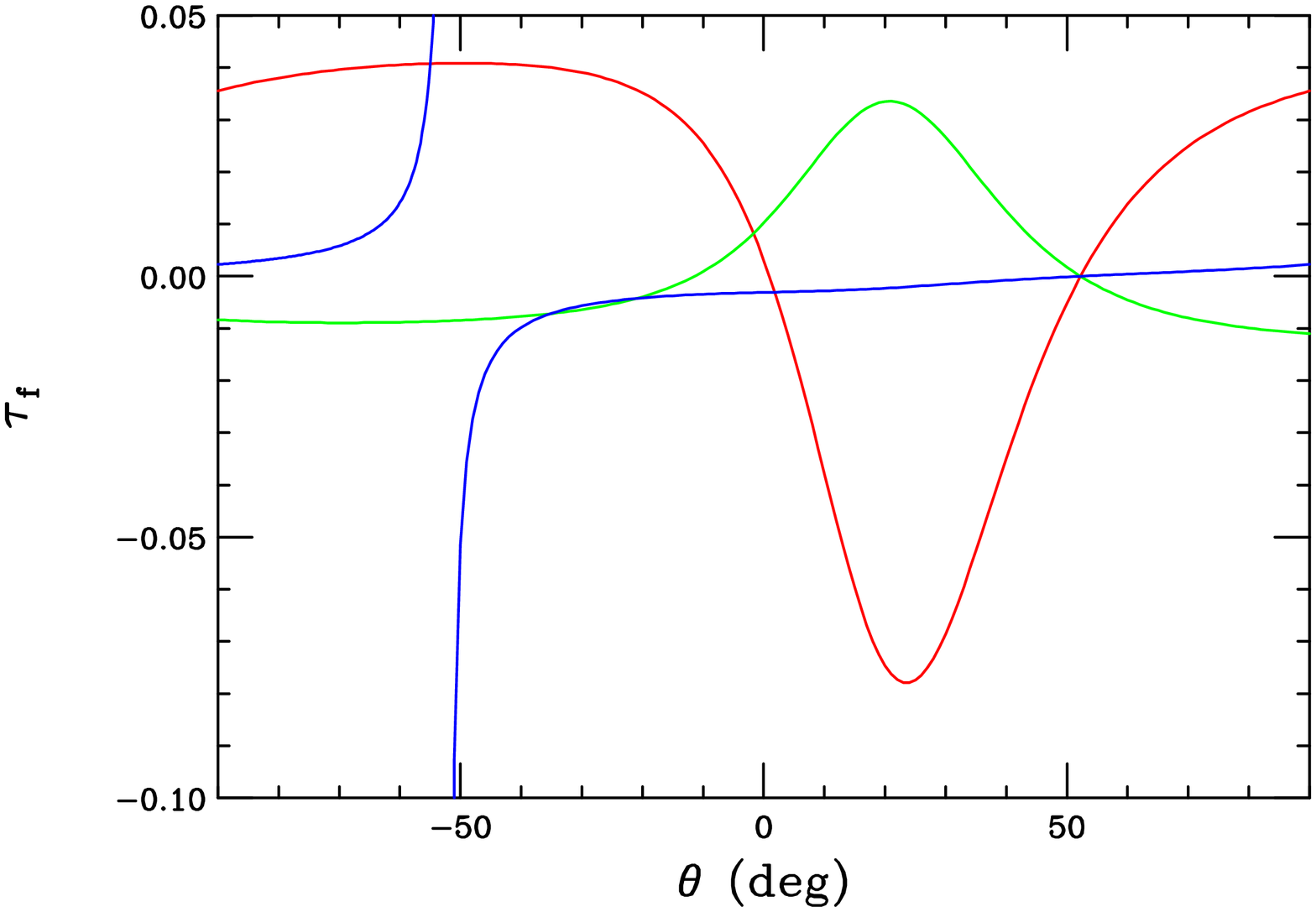}}
\vspace*{0.1cm}
\caption{Values of the observables $\lambda$(top) and $\tau_{f=l,b,c}$(bottom), as represented by the red, green and blue curves, respectively, as a function 
of the parameter $\theta$ for $Z'$ originating from $E_6$ models.}
\label{fig7}
\end{figure}

How sensitive are these observables to changes in the $Z'$ couplings and can they be used (in conjunction with the with the unpolarized observables) to obtain 
the parameters of the underlying theory? This, of course, requires a detailed study but we can get a strong indication of what may be possible by looking at a 
few examples. Fig.~\ref{fig7} shows the values of the observables $\lambda$ and $\tau_{f=l,b,c}$ for the well-known set of $Z'$ originating in $E_6$ 
models{\cite {zp}}. Note that in such models, a single parameter, $\theta$, controls all of the fermionic couplings to the $Z'$. From this figure we see that 
these transverse polarization observables are quite sensitive to the value of this parameter.

\begin{figure}[htbp]
\centerline{
\includegraphics[width=7.5cm,angle=90]{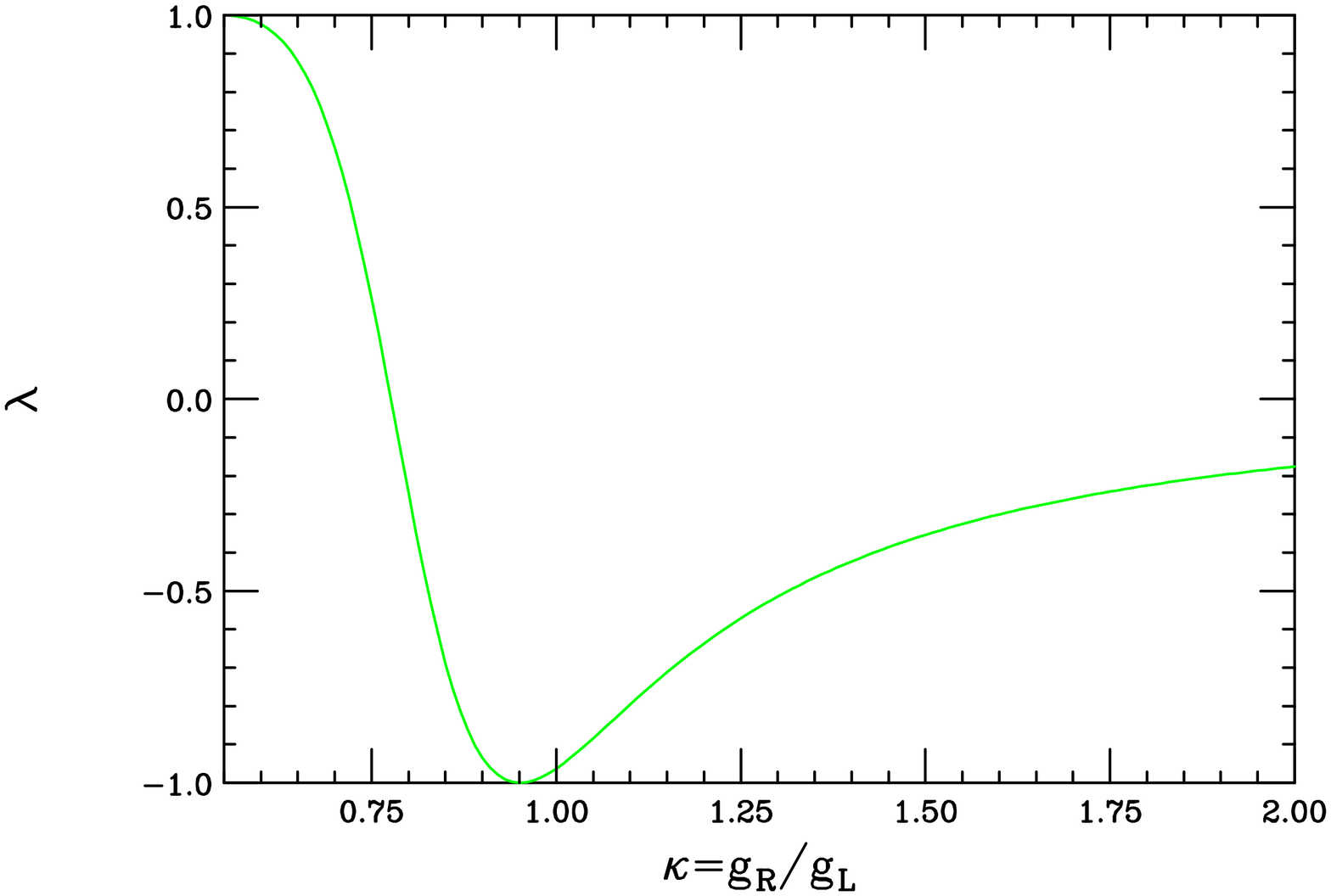}}
\vspace*{0.1cm}
\centerline{
\includegraphics[width=7.5cm,angle=90]{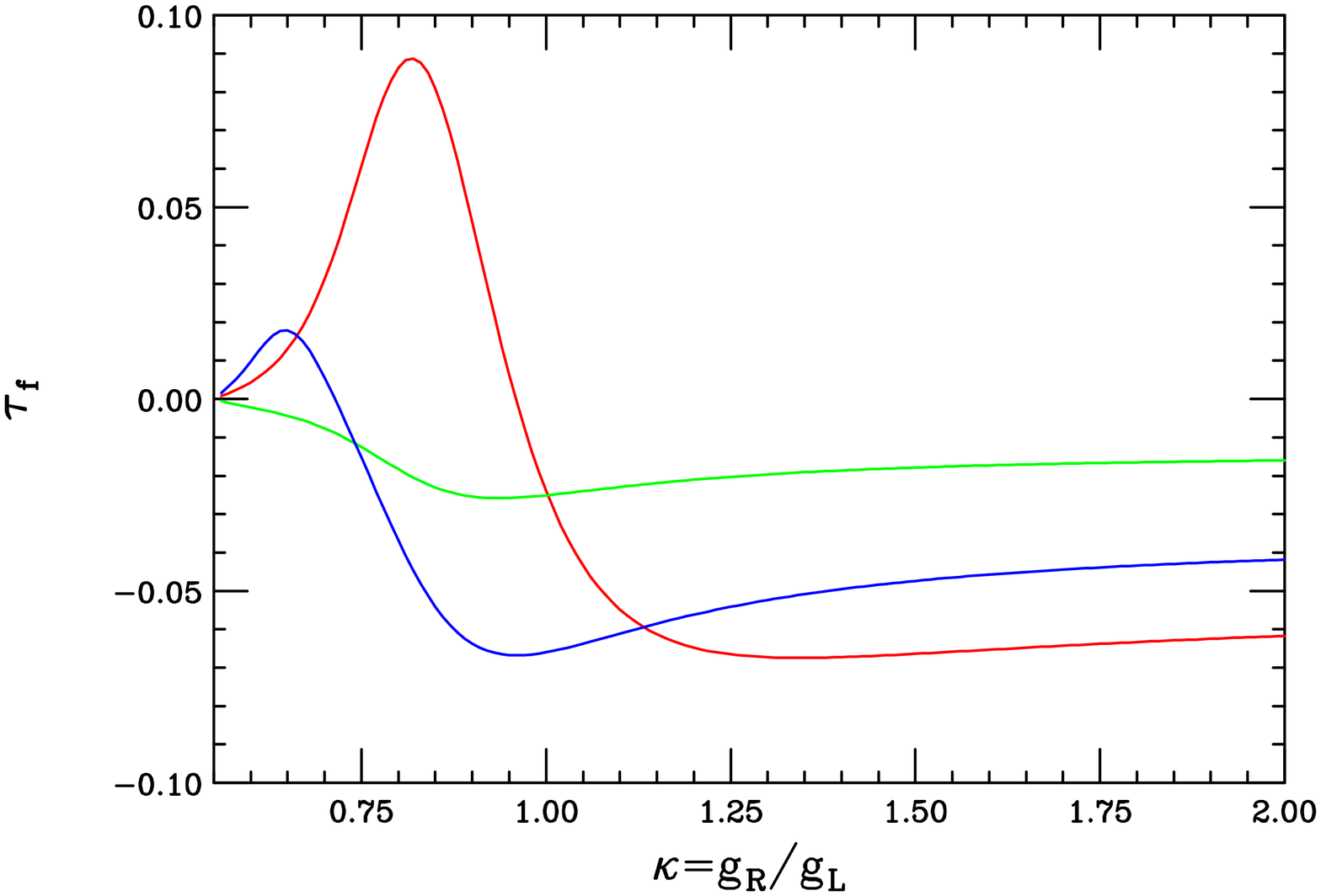}}
\vspace*{0.1cm}
\caption{Same as the previous figure but now for the Left-Right Symmetric Model as a function of the parameter $\kappa=g_R/g_L$.}
\label{fig8}
\end{figure}

As a further example to probe the coupling sensitivity of these observables, 
we consider a $Z'$ from the Left-Right Symmetric Model{\cite {zp}} where the only free parameter is the ratio of the two $SU(2)_{L,R}$ 
gauge couplings, $\kappa=g_R/g_L$. Fig.~\ref{fig8} shows the values of $\lambda$ and $\tau_{f=l,b,c}$ for this class of models. Again we see that transverse 
polarization observables are quite sensitive to the value of $\kappa$ through the various fermion couplings. Clearly, these new observables do a respectable 
job at providing substitute coupling information to that obtainable from 
$A_{LR}$ and  $A_{FB}^{pol}(f)$ when transverse polarization is available instead of longitudinal polarization. 

It is also interesting to consider what would happen if the new resonance were not a spin-1, $Z'$ but were instead a spin-2 graviton KK excitation as 
in the original version of the Randall-Sundrum model{\cite {ed,RS,us}}. In such a case the on-resonance double differential cross section 
for a massless fermion in the final state would take the generic form
\begin{equation}
{{d\sigma}\over {dz d\phi}} \sim \big(1-3z^2+4z^4\big) \big[1-P_1^T P_2^T \cos 2\phi\big]+ {{\Gamma}\over {M}}P_1^T P_2^T F(z,v,a)\sin 2\phi \,,
\end{equation}
where $F$ is a rather complex function of the couplings and $z$, 
from which we learn several things. Most importantly, the $z$-dependence of the the unpolarized cross section and that of the $\cos 2\phi$ part of the 
azimuthal distribution are seen to be identical which is somewhat reminiscent of the case of scalar particle pair production through $s$-channel spin-1 
gauge boson exchange as was discussed above. Secondly, 
we see that the value of the $\lambda$ parameter is completely fixed, \ie, it is universal and independent of the fermion flavor since gravitational couplings are 
universal in this scenario. Thirdly, as was the case for the $Z'$, a width-suppressed $\sin 2\phi$ term is again present (although we do not give its explicit 
form here via the function $F$) that depends upon the interference of the graviton with the usual SM $\gamma$ and $Z$ exchanges. Clearly this graviton KK resonance 
will be easily distinguishable from a $Z'$ at CLIC.

\section{Conclusions}

In this paper we have considered the further use of transverse polarization and the analogous azimuthal angular 
distributions as means to explore the properties of new states produced 
at the $e^+e^-$ collider CLIC running at $\sqrt s=3$ TeV. Here we have shown that ($i$) transverse polarization asymmetries can be used as a discriminator of particle 
spin; in particular, the two possibilities of smuon or UED KK-fermion pair production can be easily distinguished. ($ii$) Furthermore, we have shown that the 
general form of the azimuthal angular distribution, as measured on top of the pole of a new $Z'$-like state, can provide a powerful handle on the couplings of the 
various SM fermions to this new state in a manner analogous to observables employed in the case of longitudinal beam polarization. In both of these scenarios, further 
work will be necessary to more fully understand the power of these observables in a CLIC-like detector environment including the influence of the full SM and other 
new physics backgrounds and the effects of the significant CLIC machine-induced beamstrahlung.

\noindent{\Large\bf Acknowledgments}

The author would like to thank J.L. Hewett, R.M. Godbole, G. Moortgat-Pick and S. Riemann for various discussions related to this work.

%
\def\MPL #1 #2 #3 {Mod. Phys. Lett. {\bf#1},\ #2 (#3)}
\def\NPB #1 #2 #3 {Nucl. Phys. {\bf#1},\ #2 (#3)}
\def\PLB #1 #2 #3 {Phys. Lett. {\bf#1},\ #2 (#3)}
\def\PR #1 #2 #3 {Phys. Rep. {\bf#1},\ #2 (#3)}
\def\PRD #1 #2 #3 {Phys. Rev. {\bf#1},\ #2 (#3)}
\def\PRL #1 #2 #3 {Phys. Rev. Lett. {\bf#1},\ #2 (#3)}
\def\RMP #1 #2 #3 {Rev. Mod. Phys. {\bf#1},\ #2 (#3)}
\def\NIM #1 #2 #3 {Nuc. Inst. Meth. {\bf#1},\ #2 (#3)}
\def\ZPC #1 #2 #3 {Z. Phys. {\bf#1},\ #2 (#3)}
\def\EJPC #1 #2 #3 {E. Phys. J. {\bf#1},\ #2 (#3)}
\def\IJMP #1 #2 #3 {Int. J. Mod. Phys. {\bf#1},\ #2 (#3)}
\def\JHEP #1 #2 #3 {J. High En. Phys. {\bf#1},\ #2 (#3)}

\end{document}